\documentclass[preprint,aps]{revtex4}
\usepackage{graphicx,epsfig}

\begin{document}


\title{ Naive model from 1970th applied to CMR manganites: it seems to work }

\author{A.Vl. Andrianov}
\affiliation{Department of Physics, Moscow State University, Moscow 119991,
Russia }


\begin{abstract}

Existing experimental data for various colossal magnetoresistance manganites 
have been examined employing an ovesimplified  model that roots in 1970th. 
This model considers a classical semiconductor where conducting bands are affected 
by the strong Weiss exchange 
field that arises from the magnetic order in the substance.
The field--caused
shifts of the conducting bands results in the change in the number of 
thermally activated carriers, and
this change is presumed to be responsible for 
the resistivity dependences on temperature and magnetic field and for the CMR itself.
Employing this model we calculate this hypothetical Weiss field  from the experimental data
for various CMR manganites
employing minimal set of the adjustable parameters, namely two.
The obtained Weiss field behaves with temperature and external field similarly to 
the local magnetization, its supposed source, hence supporting the model.

\end{abstract}


\maketitle

\section{INTRODUCTION}

The importance and complexity of the colossal magnetoresistance (CMR) phenomenon in manganites, 
the incredible drop in the 
resistivity with the onset of the magnetic order and/or under external magnetic field \cite{Jin_1},
do not require to be emphasized once again. The mechanism of this fascinating phenomenon remains uncertain
despite the elaborated theoretical studies, see reviews \cite{Tokura_book}, \cite{Salamon_Jaime}.

One of the most common features of these materials is the semiconductor-like behavior of 
resistivity above the magnetic ordering temperature \cite{Salamon_Jaime}.
In \cite{FeCr2S4_us} we have revisited an apparently oversimplified model that roots in 1970th;
in addition to this model's history (see references in  \cite{FeCr2S4_us}) we have traced it
back to \cite{Oliver_PhRB} (1972) and \cite{Searle_Wang}(1969). 
Its approach is naive: 'if an object behaves as a semiconductor above the magnetic 
ordering temperature -- treat it as a true semiconductor and consider effect of the magnetic ordering.'
Nevertheless this model demonstrates a good  semiquantitative agreement with the CMR 
behavior in $FeCr_{2}S_{4}$ Ref.\cite{FeCr2S4_us}. 

In this work we are going to demonstrate that this ancient and naive model is consistent with 
the characteristic behavior of the various CMR manganites.

\section{MODEL} 

As in \cite{FeCr2S4_us}, we consider the simplest semiconductor with parabolic density 
of states in the conducting band and activation energy $\Delta$, see Fig.1a. 
The local electrons responsible for the magnetic ordering form the local magnetic 
levels (depicted by inscription and bold-dashed line) that define the Fermi energy $\epsilon_F$. 
Both $\epsilon_F$ and $\Delta$ are assumed independent on temperature and magnetic field. 
The magnetic and conducting electrons are considered separately and 
assumed to be tied only by magnetic exchange between them. 
It means that the occurrence of the magnetic order 
results in the nonzero Weiss quasi-field $H_{W}$ applied to the 
conducting electrons. We expect  $H_{W}$ to be proportional to the local
magnetization $M$, either spontaneous and/or field-induced.
External field $H$ is expected to be much smaller than $H_W$, while $H_W$ might depend strongly on $H$, 
especially in the vicinity of the Curie temperature $T_C$. 
This effective field causes the shift of the spin-up and 
spin-down conducting bands by $\mu_{B}H_{W}$ down and up respectively,
see Fig.1b. The condition $\mu_{B}H_{W}=\Delta$ corresponds to the closure of the gap and hence to the crossover 
from the semiconductor to the half-metal, sketched on Fig.1c.

The density of thermally activated carriers $n$ in this model is described by expression from Ref.\cite{FeCr2S4_us}:
\begin{equation}     
\label{trivial}
n(T,H_{W})=\frac{A}{2} (\int \limits_{0}^{\infty} 
\frac{\epsilon^{1/2}d\epsilon}{exp\frac{\epsilon+\Delta-\mu{_B}H_{W}}{kT}+1}+
\int \limits_{0}^{\infty} 
\frac{\epsilon^{1/2}d\epsilon}{exp\frac{\epsilon+\Delta+\mu{_B}H_{W}}{kT}+1})\equiv{}A*f_{W}(T,H_{W},\Delta)
\end{equation}
where $A$ is 
a numerical coefficient, first integral corresponds to spin-up band and 
second to the spin-down one. 

The typical dependences of $n$ on $H_{W}$ at different temperatures
calculated with Eq.1
are presented in Fig.2. Application of Weiss field always causes the increase in the number of carriers, i.e.
decrease in resistivity.
Note that the crossover from semiconductor to half-metal 
at $\mu_{B}H_{W}/\Delta=\pm{}1$, marked by dotted line, is not pronounced even at 
$kT$ several times smaller than $\Delta$.

In Fig.3. we present the simulation of zero-field temperature dependences of resistivity $\rho(T)$, supposedly proportional to
$1/n(T,H_{W})\propto{}1/f_{W}(T,H_{W},\Delta)$ obtained using Eq.(1). 
Weiss field $H_W$ is assumed strictly proportional to the spontaneous magnetization:
$H_{W}(T)=H_{W}(0)*M_{s}(T)/M_{s}(0)$; 
we set realistic values $T_C$=200K and $\Delta$=700K, employ the same dependence $M_s(T)$ (inset Fig.3) 
for all the dependences and vary only the strength of the Weiss field $H_W$, governed by $H_{W}(0)$ parameter. 
Its values are listed over the curves. 
The resulting temperature 
dependences look credible, resembling typical experimental dependences $\rho(T)$ for various CMR materials.

\section{ANALYSIS}

As a next step we reverse the approach and  try to obtain the suggested Weiss field 
from the experimental $\rho(T,H)$ dependences. 
The obvious problem is the contribution of the unknown carriers' mobility $\mu$ to the resistivity.
In this analysis we always set $\mu=const$ that appeared to be sufficient. Hence the final
equation is
\begin{equation}  
\label{final}
\rho(T,H)=B/f_{W}(T,H_{W},\Delta)
\end{equation}
, where $B$ is a numerical coefficient (includes mobility, carriers' effective mass and so on). 

The procedure is as follows: first we fit the high-temperature paramagnetic fragment of the experimental $\rho(T)$ 
dependence by Eq.(2) with $H_{W}$ set equal to zero, i.e. by ordinary expression for the nonmagnetic semiconductor.
Typically this fit is good enough. Obtained $\Delta$ 
and $B$ values are the only adjustable parameters of the model. 
Since that the procedure is unambiguous: we solve the Eq. (2) numerically for the 
each experimental $\rho$, $T$ and $H$ values using the same $\Delta$ 
and $B$, obtaining therefore the $H_{W}(T,H)$ dependence. We use{\it 'Mathematica 6'} package for the calculations. 
Note that there are no adjustable parameters related to magnetism at all.
According to the model this dependence is expected to follow the characteristic behavior of the local magnetization $M(T,H)$, 
the supposed source of $H_W$.

We have analyzed this way various CMR manganites using experimental data from \cite{LaYCaMnO3-x}, \cite{LaYCaMnO3-H},  \cite{SmSrMnO3-1},
\cite{Pr0.5Ca0.5CoMnO3}, \cite{La945MnO3}. All the results are in general agreement with each other. Here we present
the most conclusive outcome obtained with the data Ref.\cite{LaYCaMnO3-x}, the detailed study of $La_{0.7-x}Y_{x}Ca_{0.3}MnO_{3}$ system
where $\rho(T)$ dependences cover more than 4 decimal orders of magnitude, and Ref.\cite{LaYCaMnO3-H} with $\rho(T)$ dependences 
in $La_{0.56}Y_{0.19}Ca_{0.25}MnO_{3}$ at various magnetic fields. We also present analysis of the data Ref.\cite{SmSrMnO3-1} for
$Sm_{0.55}Sr_{0.45}MnO_{3}$ and Ref.\cite{Pr0.5Ca0.5CoMnO3} for $Pr_{0.5}Ca_{0.5}Mn_{0.95}Co_{0.05}O_{3}$ 
with the step-like magnetic phase transition. 

Fig.4 presents $H_{W}(T)$ dependences  for $La_{0.7-x}Y_{x}Ca_{0.3}MnO_{3}$ ceramics at various $x$, calculated 
from data Fig.1 Ref.\cite{LaYCaMnO3-x} with activation energy $\Delta$=1400K for all the data. The $H_{W}(T)$ dependence for $x=0$ 
(dashed line, guide for the eye)
resembles nicely typical $M_{s}(T)$ behavior (compare inset Fig.3.) All the dependences for $x\ne{}0$ follow the same dashed line
scaled properly vertically and horizontally. The deflection at $H_{W}\to{}0$ is due to the broadening of the 'knee' in $\rho(T)$ near $T_C$ 
with $x$ increase, see original Fig.1 Ref.\cite{LaYCaMnO3-x}. Therefore not only 
obtained $H_{W}(T)$ dependences are credible, but also
strength of the Weiss field declines slightly but regularly with $x$ increase, as could be expected.

The $H_{W}(T)$ isofield dependences for polycrystalline $La_{0.56}Y_{0.19}Ca_{0.25}MnO_{3}$  are presented on Fig.5; 
data from Fig.4. Ref.\cite{LaYCaMnO3-H}. 
We obtain a regular behavior that is fully consistent with the $M(T)$ dependences in
the typical ferromagnet under magnetic field: increase in $H$ always causes increase in $H_W$, $H_W$ is almost field independent 
well below $T_C$ but depends on $H$ strongly in the vicinity of $T_C$; $H_W$ depends on $H$ roughly linearly well above $T_C$. 
Moreover, the dependence for $H=$5T even follows the Curie-Weiss dependence at $T>T_C$, see inset Fig.5. 
The similar result have been obtained for $La_{0.945}Mn_{0.945}O_{3}$, Fig.7. Ref.\cite{La945MnO3} (not presented).

Ref.\cite{LaYCaMnO3-H} allows to compare the calculated Weiss field with the experimental magnetization in 
$Pr_{0.5}Ca_{0.5}Mn_{0.95}Co_{0.05}O_{3}$ ceramics where magnetic transition is step-like.
On Fig.6 we present the zero-field $H_{W}(T)$ dependence (left scale) obtained with Eq.(2) from 
Fig.2 Ref.\cite{LaYCaMnO3-H} (curve 'Co 5\%, H=0') with the experimental FC magnetization $M(T)$ 
measured in field $H=$1.45T (right scale). These two dependences behave in a quite similar way.
Note that the applied field shifts the magnetic transition in this substance to higher temperatures \cite{LaYCaMnO3-H}, that is 
why the $M(T)$ dependence is shifted right -- otherwise it would match almost exactly.

Data for $Sm_{0.55}Sr_{0.45}MnO_{3}$ single crystal from Fig.2 Ref.\cite{SmSrMnO3-1} also allow direct comparison 
of the calculated $H_W$ with the experimentally measured magnetization in various magnetic fields. The magnetic phase 
transition in this substance is first-order and the associated drop in resistivity exceeds 3 orders of magnitude.
On Fig.7 we compare the $H_W{}(T)$ isofield dependences calculated from resistivity data Fig.2(b) Ref.\cite{SmSrMnO3-1} by 
the same procedure as before (left scale) and the isofield imagnetization $M(T)$ Fig.2(a) Ref.\cite{SmSrMnO3-1} (right scale). 
The $\Delta$=1200K value, obtained from $H=$0 dependence, is the same for all the data. Vertical scales adjusted to obtain the best match.

We see that the calculated $H_W{}(T)$ dependences for $H=$3T and $H=$4T follow nicely the experimental $M(T)$ curves 
at the same field values. The agreement is surprisingly good taking into account the uncertainty in carriers' mobility. It looks like
the change in carriers' concentration under magnetic ordering is so huge that the simultaneous change in mobility is 
negligible in comparison. 

Therefore all the examined experimental data appear to be consistent with the model. The hypothetical Weiss field 
always behaves in the same style as the local magnetization -- its suggested source.

\section{DISCUSSION}

Certainly the demonstrated agreement and consistency, while undeniable, are insufficient to admit this model.

The obvious problem of the model is the value of the calculated $\mu_B{}H_W{}>{}10^3$K that exceeds $T_C$
several times. Taken literally it means that the exchange between a thermally activated carrier 
and a magnetic spin is several times stronger than
the exchange between these spins responsible for the very magnetic ordering.
It favors the formation of magnetic polarons \cite{Kasuya_polaron} instead of the lone thermally activated carriers. 

Moreover the suggested gap $\Delta$ is most likely a pseudogap \cite{Saitoh_pseudogap}.

We even cannot rule out that the model provides reasonable results only due to some coincidence.

Nevertheless we hope that the demonstrated consistency with the most magnetotransport features
of the various CMR manganites as well as the transparency of the model are convincing enough 
to draw attention to this approach.

\section{CONCLUSION}

In conclusion, we consider a classical semiconductor affected
by the hypothetical Weiss exchange field that arises 
from the magnetic order. This is likely the most straightforward  approach to the 
rearrangement of the band structure with the onset of magnetic order.
The results obtained in this phenomenological model give
a credible description of the resistive and magnetoresistive properties 
of CMR manganites. The
Weiss field refined from the experimental data behaves with temperature and external magnetic 
field in the same style as the local magnetization, the suggested source of this Weiss field.
Therefore this oversimplified and naive model nevertheless deserves attention.


\newpage

\begin{figure} [tbh]  
\resizebox{0.9\columnwidth}{!}{\includegraphics*{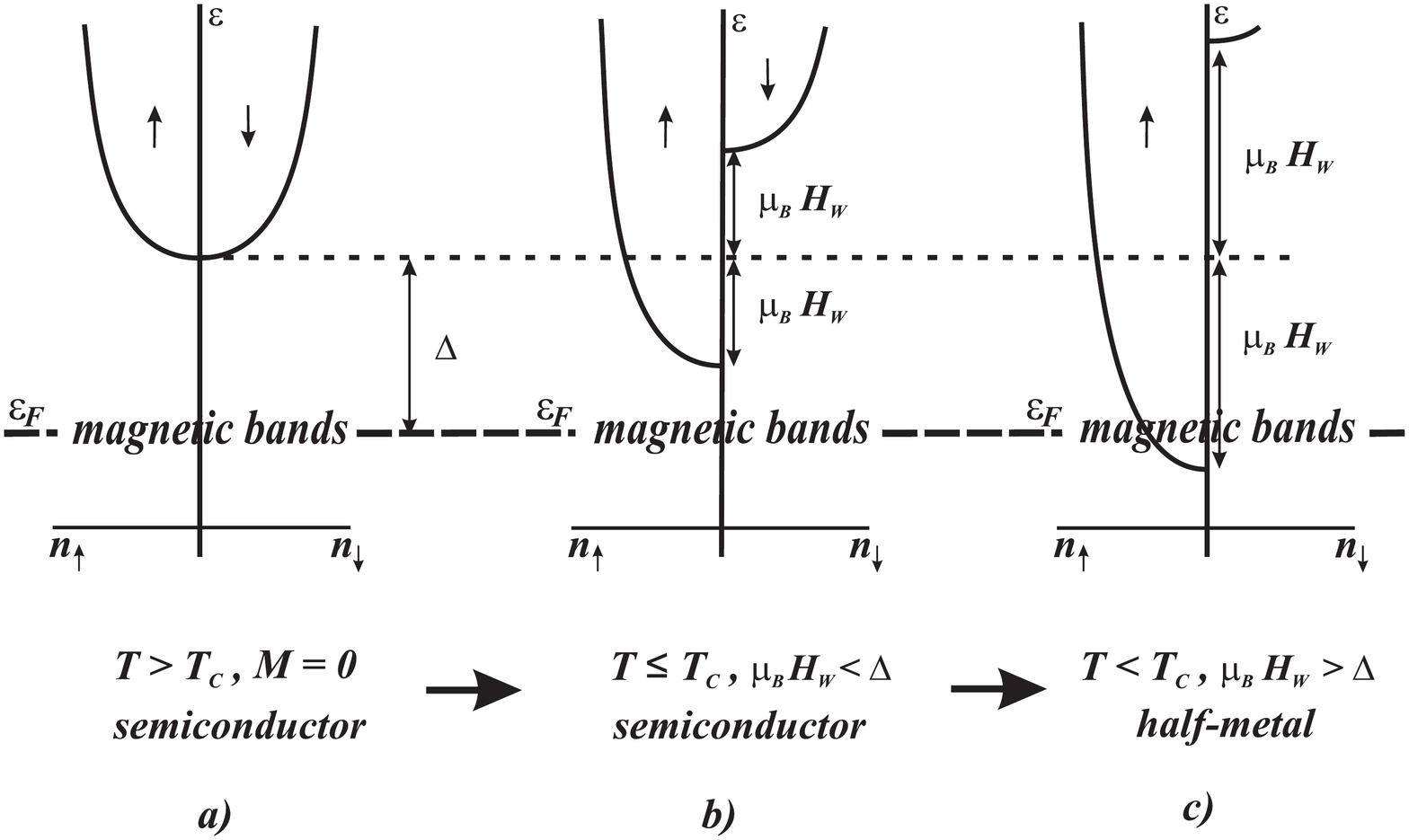}}
\caption{
a) The basic model in the absence of the magnetic field and magnetic order: the density
of spin-up $n_\uparrow$ and spin-down  $n_\downarrow$ states as a function of energy $\epsilon$. 
$\Delta$ is the activation energy, the conducting band 
(at the top) is assumed parabolic.
Narrow magnetic bands, depicted by inscription, are
responsible for the magnetic order, they also define the Fermi energy $\epsilon_F$ (
dashed horizontal line). 
b) The band structure after occurrence of the magnetization with temperature 
and/or external magnetic field.
A strong Weiss field $H_{W}$ arises from the magnetic order. It shifts the conducting 
bands by $\mu_{B}H_{W}$ down and up respectively. 
The change in the magnetic bands associated with the magnetic ordering neglected, 
$\Delta$ and $\epsilon_F$ are assumed constants.
c) The crossover from the semiconductor to the half-metal with $H_{W}$ increase.
}
\end{figure}

\begin{figure} [tbh]   
\resizebox{0.9\columnwidth}{!}{\includegraphics*{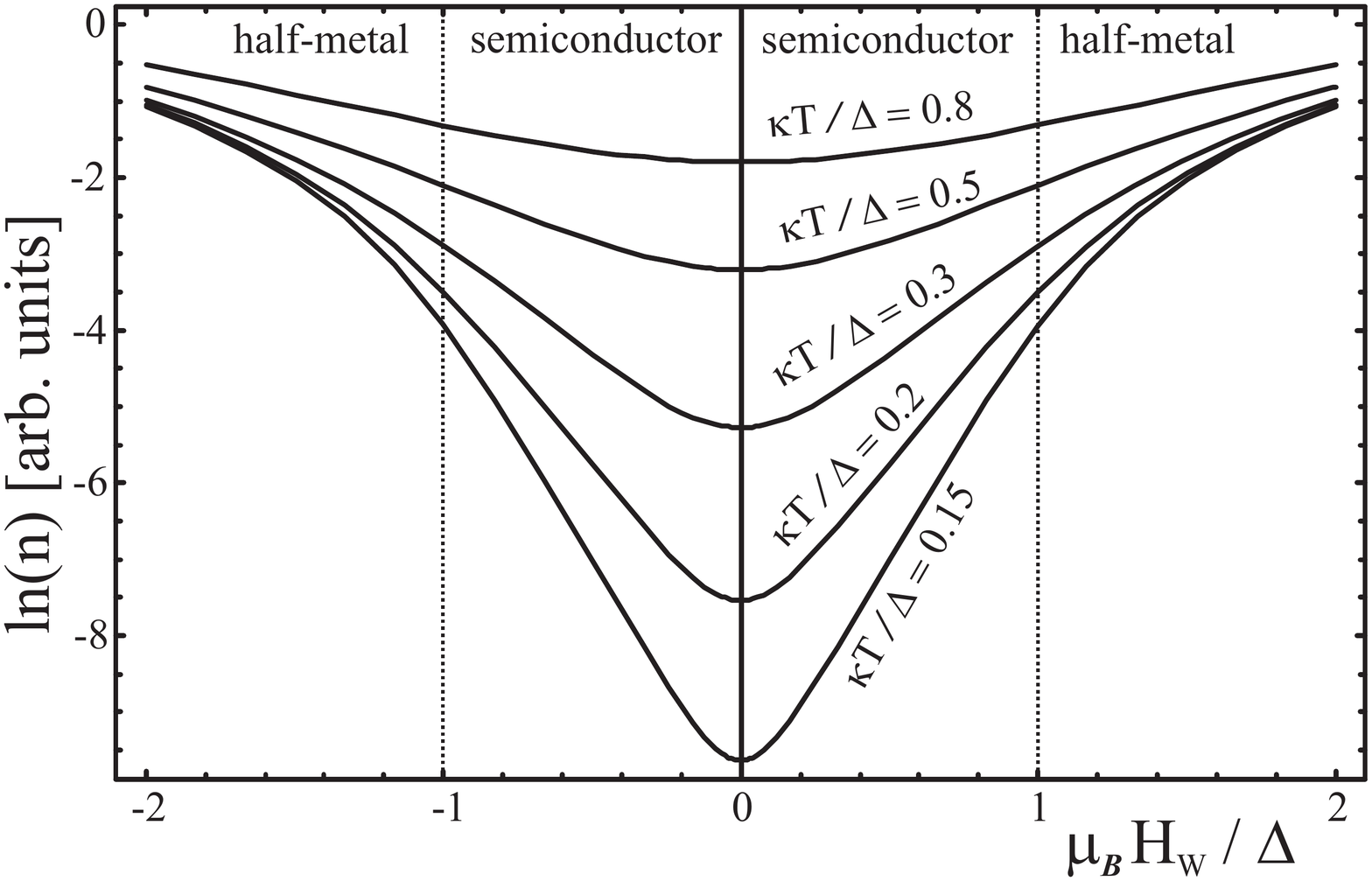}}
\caption{
The density of carriers $n$ as a function of the Weiss field $H_{W}$ at 
various temperatures (arbitrary units, semilogarithmic scale) calculated using Eq.(1). 
Dotted vertical lines
mark crossover from the semiconductor to the half-metal.
}
\end{figure}

\begin{figure}   
\resizebox{0.9\columnwidth}{!}{\includegraphics*{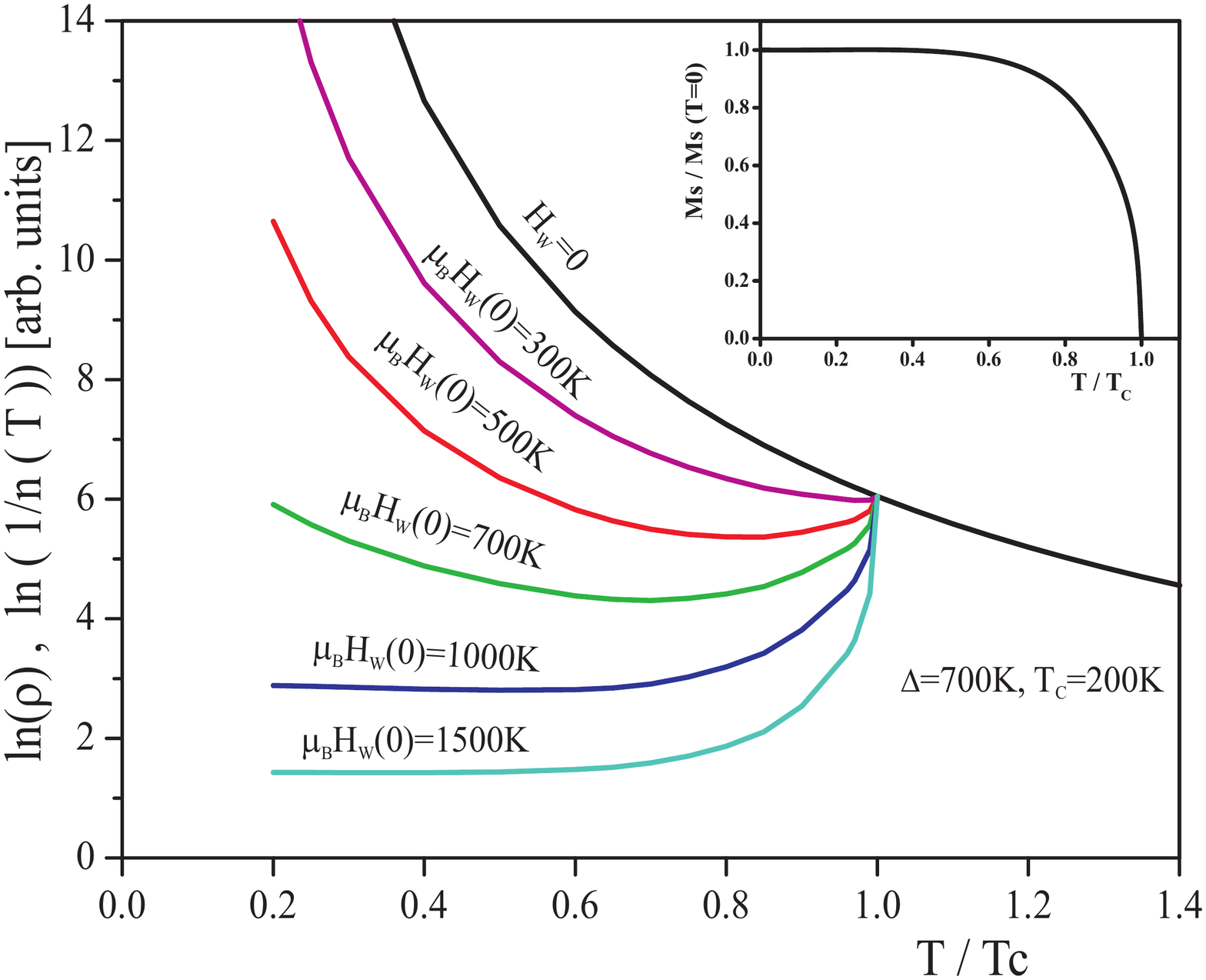}}
\caption{
(Color online.) Simulations of the zero-field resistivity as a 
function of the reduced temperature (logarithmic scale, arbitrary units). 
Weiss field $H_{W}$ assumed proportional to 
the spontaneous magnetization $M_s$, lines are the results of calculation by Eq.(2). 
Inset: the $M_{s}(T)$ dependence used for calculations.
Values of the Weiss field at $T\to{}0$ listed at the curves, $\Delta=$700K, $T_C=$200K.
}
\end{figure}

\begin{figure}   
\resizebox{0.9\columnwidth}{!}{\includegraphics*{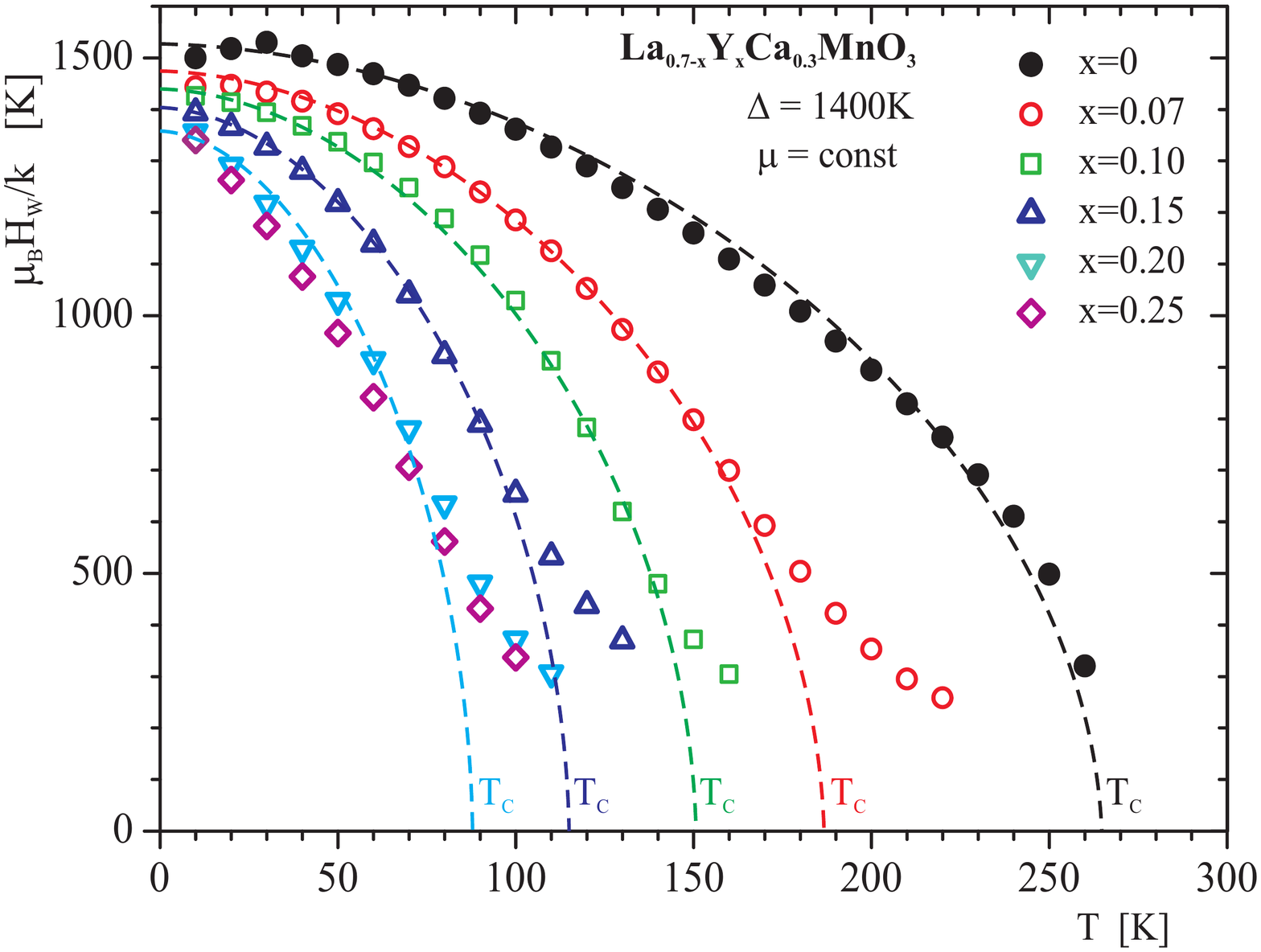}}
\caption{
(Color online.) 
Zero-field dependences of the calculated Weiss field on temperature $H_{W}(T)$ for
$La_{0.7-x}Y_{x}Ca_{0.3}MnO_{3}$ system with $x$ varied.
Obtained with Eq.(2) from the experimental dependences Fig.1a Ref.\cite{LaYCaMnO3-x}. 
The $x$ values listed on the plot. 
The $\Delta=1400$K value is an average over all the $\rho(T)$ fits by Eq.(2) in the paramagnetic state.
Dashed lines are the same 'guide to the eye' line
rescaled vertically and horizontally for each $x$ value. 
}
\end{figure}

\begin{figure}   
\resizebox{0.9\columnwidth}{!}{\includegraphics*{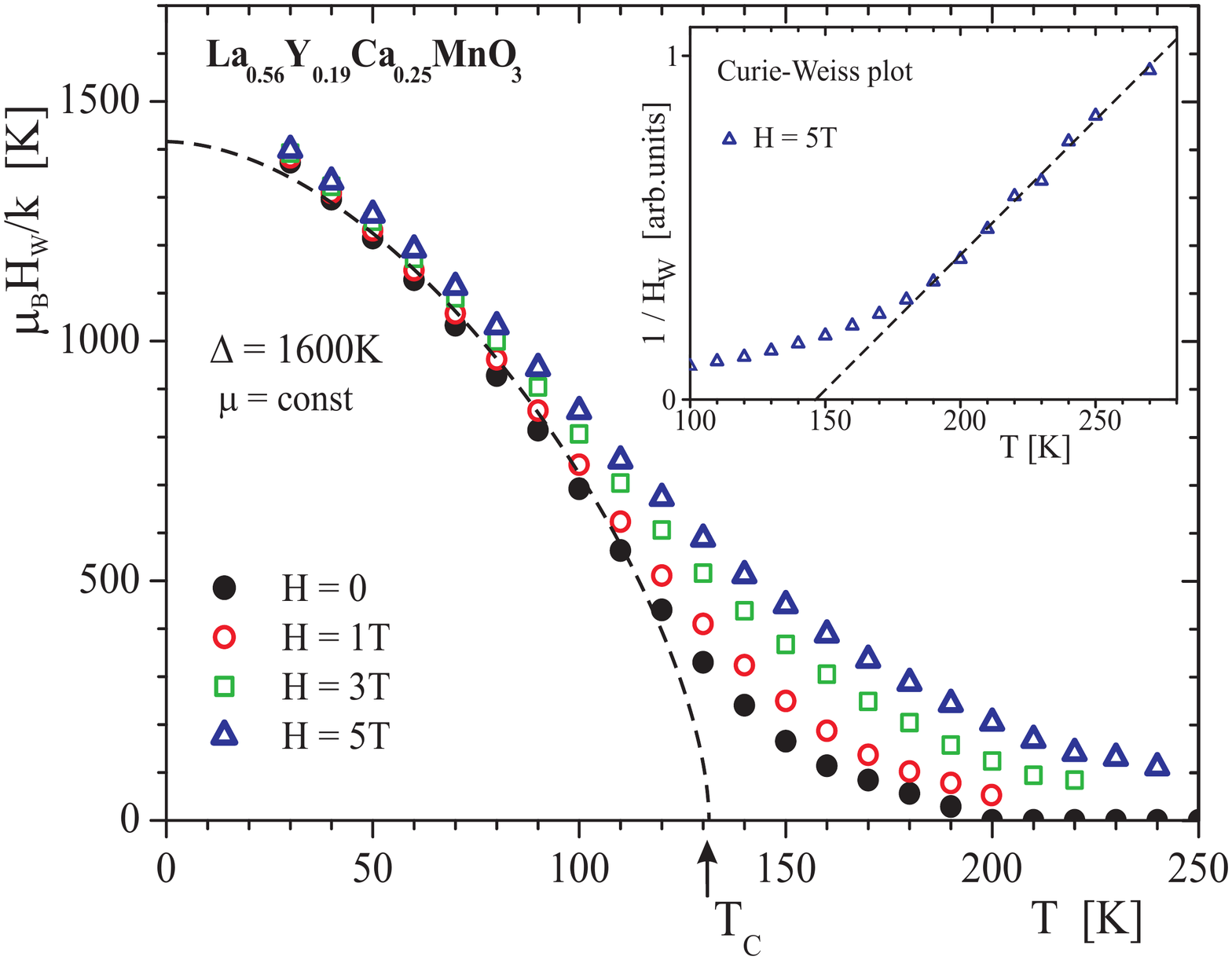}}
\caption{
(Color online.) 
Dependences of the calculated Weiss field on temperature $H_{W}(T)$ at various external fields $H$ for
$La_{0.56}Y_{0.19}Ca_{0.25}MnO_{3}$.
Obtained with Eq.(2) from experimental dependences Fig.4 Ref.\cite{LaYCaMnO3-H}. 
The $H$ values listed on the plot. Dashed line is the guide to the eye.
Inset: Curie-Weiss plot for the $H=$5T dependence.
}
\end{figure}

\begin{figure}   
\resizebox{0.9\columnwidth}{!}{\includegraphics*{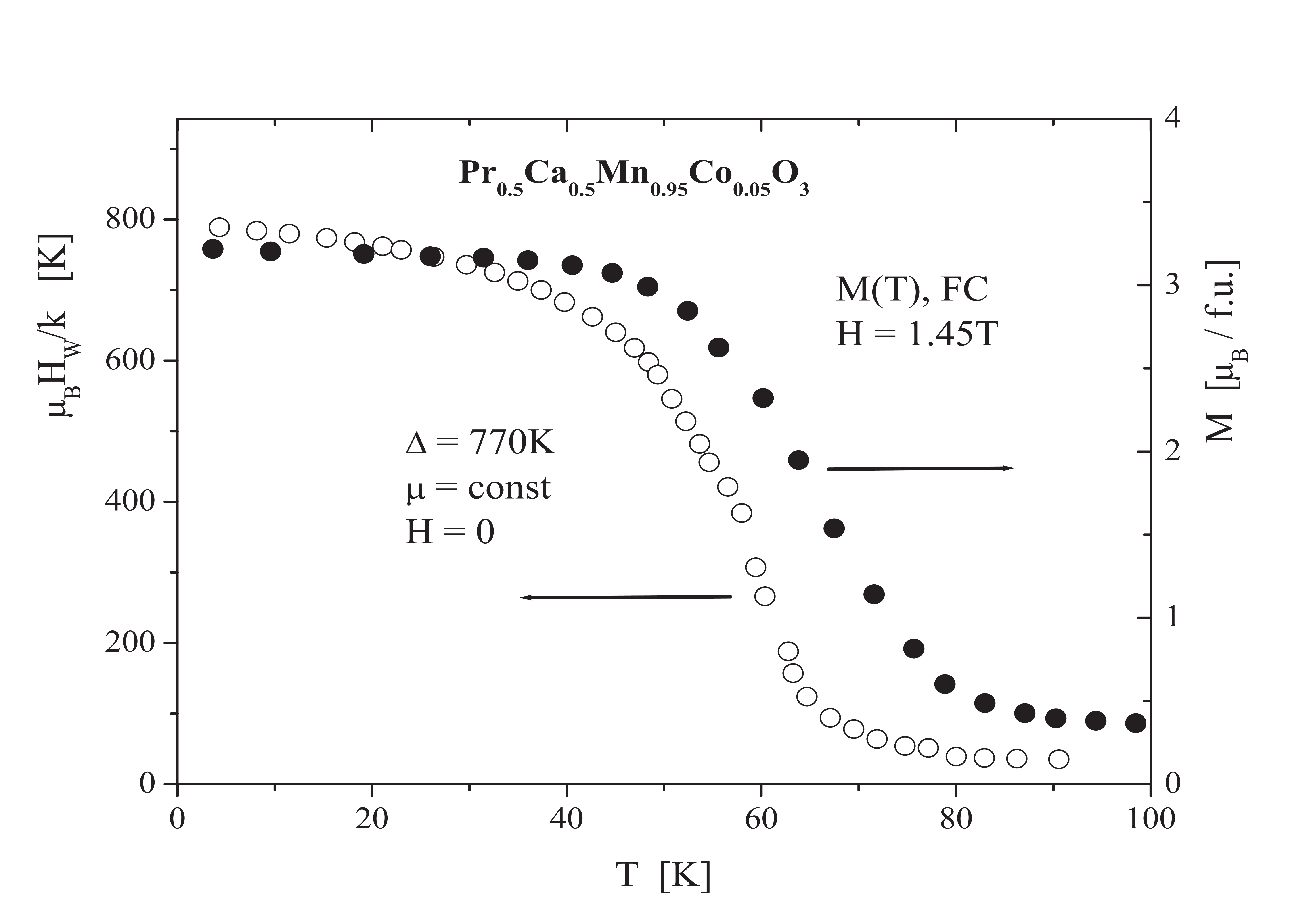}}
\caption{
(Color online.) 
Zero-field dependence of the calculated Weiss field on temperature $H_{W}(T)$ for 
$Pr_{0.5}Ca_{0.5}Mn_{0.95}Co_{0.05}O_{3}$ obtained with Eq.(2) from 
experimental dependence Fig.2 Ref.\cite{LaYCaMnO3-H} (open points, left scale) and
field-cooling magnetization $M(T)$ in $H=$1.45T from Fig.1(a) Ref.\cite{LaYCaMnO3-H}
(closed points, right scale.) Vertical scales adjusted to obtain the best match, magnetic 
field shifts transition to higher temperatures.
}
\end{figure}

\begin{figure}   
\resizebox{0.9\columnwidth}{!}{\includegraphics*{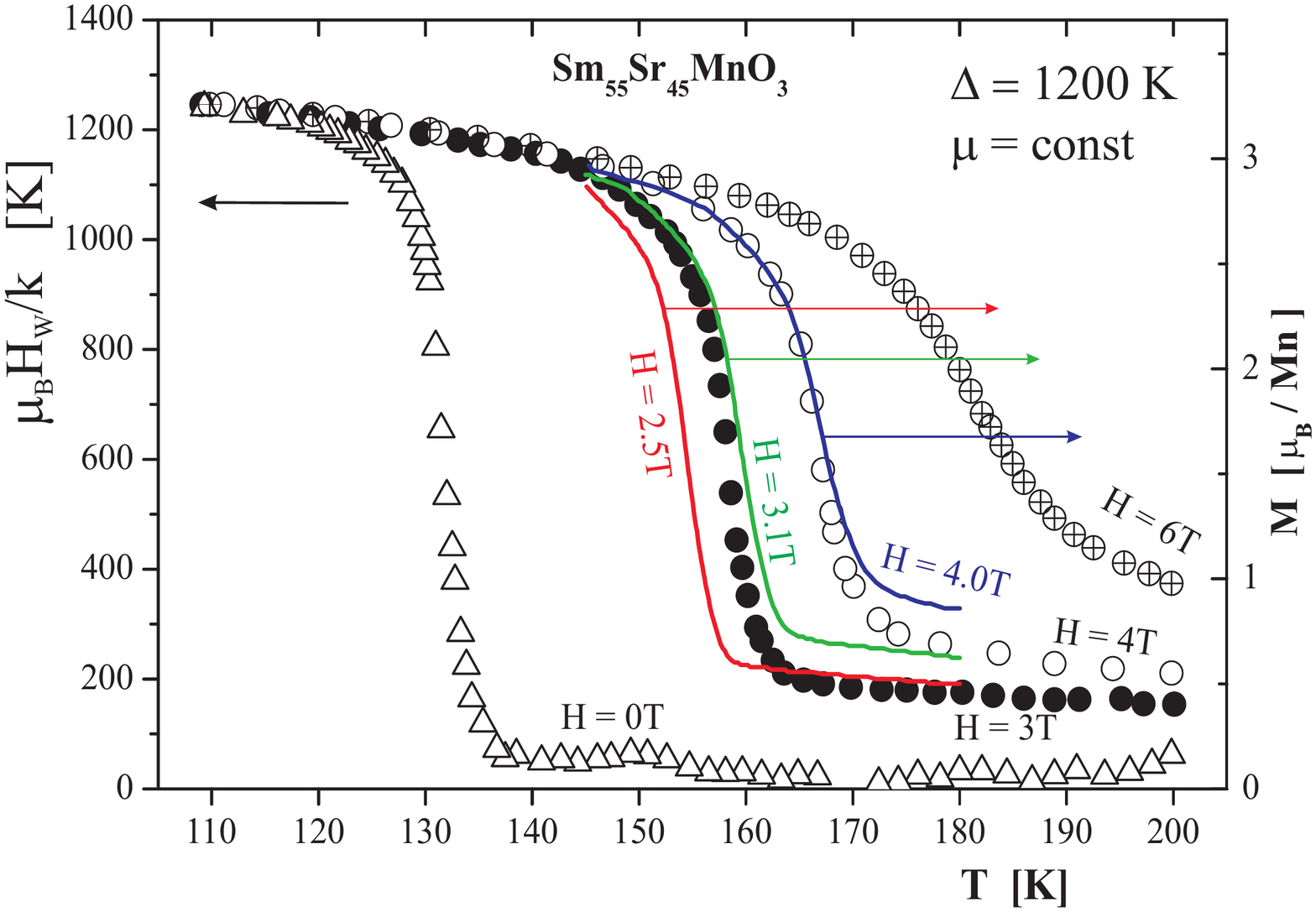}}
\caption{
(Color online.) 
Isofield dependences of the calculated Weiss field $H_{W}(T)$ (points, left scale) obtained 
with Eq.(2) from experimental dependences Fig.2b) Ref.\cite{SmSrMnO3-1} 
and the experimental magnetization $M(T)$ (lines, right scale) Fig.2a) Ref.\cite{SmSrMnO3-1}
on temperature for  $Sm_{0.55}Sr_{0.45}MnO_{3}$.
The $H$ values listed on the plot.  Vertical scales adjusted to obtain the best match.
}
\end{figure}

\end{document}